\documentclass[useAMS,usenatbib]{mn2e}
\usepackage[hyphens]{url}
\usepackage[draft]{hyperref}
\usepackage[hyphenbreaks]{breakurl}
\usepackage{longtable}
\usepackage{amsmath,amssymb,multirow,dcolumn,fancyhdr,charter,graphicx}
\usepackage{graphics}
\usepackage{xspace}
\usepackage{color,ulem,epstopdf}

\usepackage{tikz}

\newcommand{\reb}{{\sc \tt REBOUND}\xspace}
\newcommand{\whfast}{{\sc \tt WHFAST}\xspace}

\newcommand{\emcee}{{\sc \tt EMCEE}\xspace}
\newcommand{\Lagr}{\mathcal{L}}
\newcommand{\R}{R2012\xspace}

\title[Analysis of HD155358]{Resonant structure, formation and stability of the planetary system HD155358}
\author[Silburt \& Rein]{Ari Silburt$^{1,2}$, Hanno Rein$^{2,1}$\\
$^1$ Department of Astronomy and Astrophysics, University of Toronto, Toronto, Ontario, M5S 3H4, Canada\\
$^2$ Department of Physical and Environmental Sciences, University of Toronto at Scarborough, Toronto, Ontario M1C 1A4, Canada\\
}
\date{Draft version: \today}

\begin{document}
\maketitle

\begin{abstract} 
Two Jovian-sized planets are orbiting the star HD155358 near exact mean motion resonance (MMR) commensurability. 
In this work we re-analyze the radial velocity~(RV) data previously collected by \citet{Robertson2012}.
Using a Bayesian framework we construct two models -- one that includes and one that excludes gravitational planet-planet interactions (PPI).
We find that the orbital parameters from our PPI and noPPI models differ by up to $2\sigma$, with our noPPI model being statistically consistent with previous results.
In addition, our new PPI model strongly favours the planets being in MMR while our noPPI model strongly disfavours MMR.
We conduct a stability analysis by drawing samples from our PPI model's posterior distribution and simulating them for $10^9$ years, finding that our best-fit values land firmly in a stable region of parameter space.

We explore a series of formation models that migrate the planets into their observed MMR.
We then use these models to directly fit to the observed RV data, where each model is uniquely parameterized by only three constants describing its migration history.
Using a Bayesian framework we find that a number of migration models fit the RV data surprisingly well, with some migration parameters being ruled out.

Our analysis shows that planet-planet interactions are important to take into account when modelling observations of multi-planetary systems.
The additional information that one can gain from interacting models can help constrain planet migration parameters.
\end{abstract}

\begin{keywords}
planets and satellites: dynamical evolution and stability, planets and satellites: formation, planets and satellites: fundamental parameters, methods: data analysis, methods: numerical, methods: statistical. 
\end{keywords}


\section{Introduction}
\label{sec:intro}
The past decade has marked an explosion in the number of discovered multi-planet systems, with almost 600 known to date \citep{NASAEA}. 
Jovian planets have been detected in these exoplanetary systems, and the range of conditions under which they can form is uncertain.
For example, many Jovian planets are observed inside the snowline of their host star \citep[e.g.][]{Hayashi1981}, and it is unclear whether they formed in-situ \citep{Boley2016, Huang2016, Batygin2016} or beyond the snowline and migrated in afterwards \citep{Mayor1995, Lin1996, Pollack1996}.
For systems near mean motion resonance (MMR), planetary migration offers a natural formation mechanism \citep[e.g.][]{Lee2002, Rein2009}.
Since roughly $30\%$ of Jovian-sized planets with a detected planet companion are near a first-order MMR \citep{NASAEA}, at least a substantial fraction of planetary systems are likely to have formed via migration. 

The star HD155358 hosts two observed Jovian-sized planets orbiting inside the snowline near a MMR. 
The star has a mass of $0.92~M_{\odot}$, and the planets have orbital periods $P_1 = 194$ and $P_2 = 392$~days, respectively, about 2\% away from the exact 2:1 commensurability \citep[][hereafter \R]{Robertson2012}. 
In addition, HD155358 is also among the lowest metallicity stars known to host Jovian-sized planets \citep{Cochran2007}.
Since the presence of MMR can provide additional constraints on formation and evolution, HD155358 provides an opportunity to better understand the formation of Jovian planets in exoplanetary systems.

The HD155358 system has been previously studied by many scientists \citep{Cochran2007,Fuhrmann2008,Robertson2012,Andre2016}.
Of particular interest is the work by \R, who updated the orbital parameters initially reported by \citet{Cochran2007} after collecting additional radial velocity (RV) observations.
In \R the orbital properties were derived from \textsc{GaussFit\-} \citep{Jefferys1988} and \textsc{SYSTEMIC} \citep{Meschiari2009}, and planet-planet interactions were not included in their analysis (Robertson 2016, private communication). 
For well-separated orbits this assumption is valid and simplifies the analysis.
However, near a MMR this approximation can lead to different orbital solutions and therefore different formation constraints and evolutionary predictions. 
In particular, without planet-planet interactions one cannot draw any conclusion as to whether the system is in resonance or not. 

In this work we re-analyze HD155358 using the RV observations collected by \R.
In Sect.~\ref{sec:orb}, we derive new orbital parameters using a Bayesian framework coupled to direct N-body integrations and assess the probability that the system is in resonance.
In Sect.~\ref{sec:form}, we perform simulations constraining the migration history, and in Sect.~\ref{sec:stability} conduct a stability analysis. 
We conclude with a discussion in Sect.~\ref{sec:conc}.

\section{Best-Fit Parameters and Resonance Analysis}
\label{sec:orb}
\subsection{Methods}
\label{sec:Fit}
The RV data used for this analysis comes from Table 2 of \R. 
We model this system using \reb \citep{Rein2012}, an open-source N-body code for simulating problems in planetary dynamics.
Using \reb we can extract the motion of the central star over the course of our simulation, comparing it directly to the RV data. 

In this work we explore two different models -- one that includes the planet-planet gravitational interactions, abbreviated as PPI model, and one that excludes planet-planet gravitational interactions, abbreviated as noPPI model. 
This is achieved in \reb by initializing each planet as either a massive body (PPI model) or a semi-active particle (noPPI model).
Semi-active particles can gravitationally interact with massive bodies (e.g. the central star), but cannot interact with other semi-active particles. 
For reference, \R does not include the gravitational interactions between the planets, and is analogous to our noPPI model.  

We construct our models using a Bayesian framework, where the quantity of interest is the posterior probability density function. 
Benefits of conducting an analysis within a Bayesian framework include marginalization over nuisance parameters, preservation of correlations between parameters, and a natural Occam's razor when comparing models \citep[see e.g.][for a full discussion]{Gregory2005}.
Using Bayes' theorem we calculate the (unnormalized) posterior probability according to:
\begin{equation*}
p(\theta | D) \propto p(D | \theta) p(\theta),
\label{eq:Bayes}
\end{equation*}
where $\theta$ are the parameters in our model, $D$ is our data, $p(\theta)$ is our prior probability on $\theta$, $p(D | \theta)$ is the probability of $D$ given $\theta$ (i.e. the likelihood function), and $p(\theta | D)$ is our posterior probability.
As with most models that employ Bayes' theorem, the posterior probability cannot be analytically computed, requiring the use of numerical methods for approximating it. 
We use \emcee, an open-source affine-invariant Markov chain Monte Carlo (MCMC) routine  \citep{Foreman-Mackey2013}.  

For our models we assume that the system is co-planar, and use the known value of stellar mass, $m_* = 0.92M_{\odot}$.
For each planet we have the following parameters: the reduced mass, $m\sin(i)$, the semi-major axis, $a$, the eccentricity, $e$, the argument of periapsis, $\omega$, and the mean anomaly $M$. 
To avoid singularities for $\omega$ (which is ill-defined when $e\sim0$), we fit $h=e\sin(\omega)$ and $k=e\cos(\omega)$ instead of $e$ and $\omega$.
We also add an offset parameter $\gamma$ to account for any stellar drift along the line of sight, a jitter parameter $J$ to account for any stellar noise and unreported instrument variability, and a viewing angle parameter sin$(i)$.
This yields a total of 13 parameters to be sampled by the MCMC. 

We initialize our MCMC chain with values similar to the best fit values of \R to speed up convergence, and use the following priors on our parameters: $0.4M_J<m_1\textrm{sin}(i) < 2M_J$, $0.4M_J<m_2\textrm{sin}(i) < 2M_J$, $0.2 \textrm{AU}<a_1< 0.8\textrm{AU}$, $0.8\textrm{AU}<a_2< 1.4\textrm{AU}$, $1<h_1, h_2, k_1, k_2<-1$, $0<\omega_1, \omega_2, M_1, M_2<2\pi$, $-40\textrm{m/s}<\gamma<40\textrm{m/s}$, and $0\textrm{m/s}<J^2<50\textrm{m/s}$.
Our MCMC chain consists of an initial burnin phase of 1000 steps with 400 walkers, after which the MCMC walkers are resampled near the best solution and run for 5000 steps.
In \reb we use the \whfast integrator \citep{Rein2015b} with a timestep of $P_1/200$, leading to a bounded relative energy error of $<10^{-7}$.


\subsection{Best Fit Parameters}
\label{sec:Results}
\begin{figure*}
\includegraphics[trim=4.4cm 8cm 4.5cm 0cm, width=\textwidth]{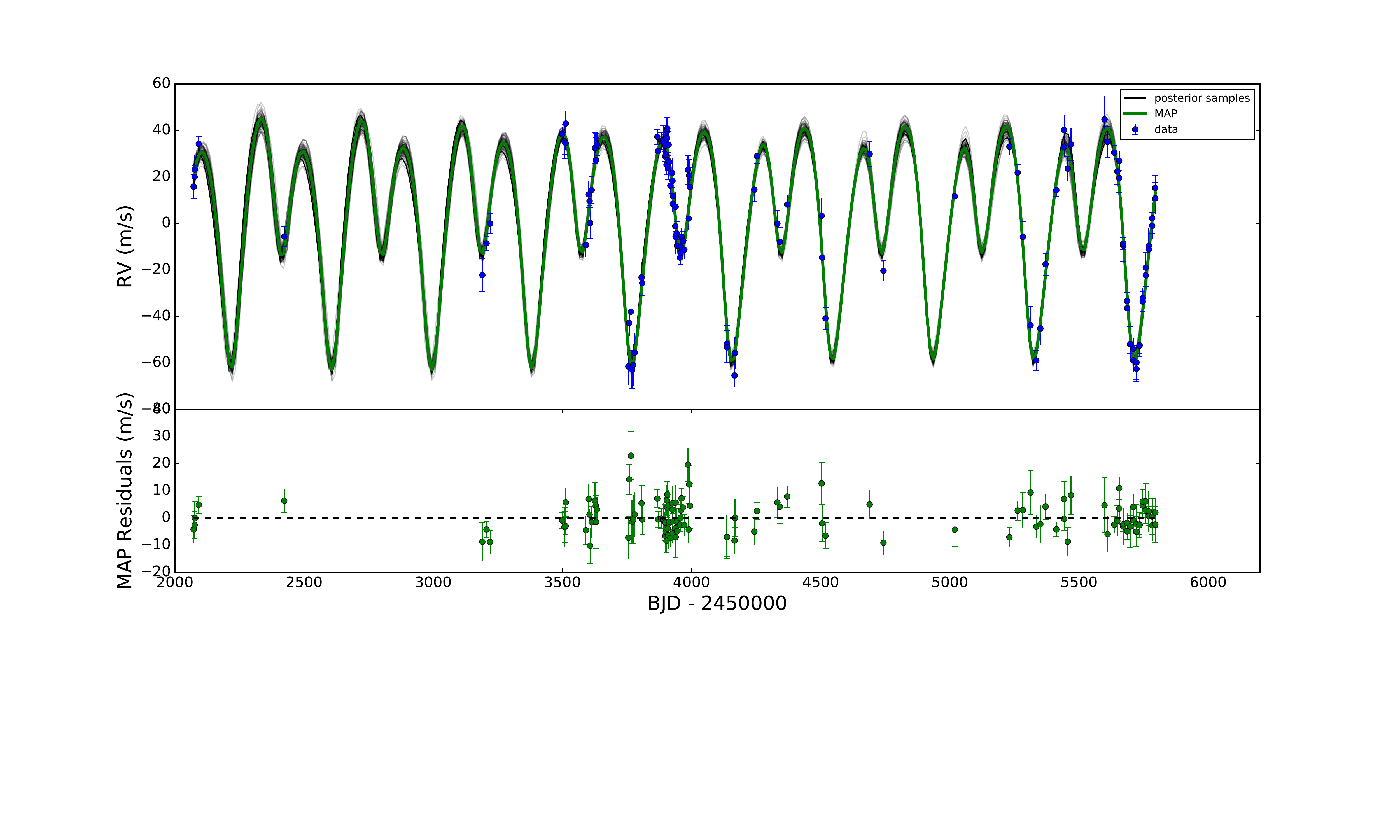}
\caption{RV data and PPI model fit. 
Top panel shows the RV data points, MCMC MAP value (green), and 80 randomly plotted samples from the posterior (black). 
Bottom panel shows the residuals for the MAP fit to the RV data. 
 }
\label{fig:MCMC}
\end{figure*}

Our maximum a-posteriori (MAP) values for each parameter are displayed in Table~\ref{tab:MCMC}. 
For ease of comparison between our results and \R, we convert back into $e$ and $\omega$ variables from $h$ and $k$ variables.
Since we find $\sin(i)$ unconstrained in our MCMC analysis (see Figure~\ref{fig:stability}) it is not listed in Table~\ref{tab:MCMC}.
Parameters with subscript $1$ relate to the inner planet, while parameters with subscript $2$ relate to the outer planet. 
A random sample of 2000 draws from our PPI model's posterior distribution is shown in Figure~\ref{fig:stability}.

From Table~\ref{tab:MCMC} there are a few statistically significant differences between our PPI and noPPI models. 
In particular, $e_2$ is inconsistent at the $1$-$\sigma$ level, while $m_1\sin(i)$, $\omega_1$ and $M_1$ are inconsistent at the $2$-$\sigma$ level.
Furthermore, every parameter in our noPPI model is statistically consistent with \R.

\begin{table}
\begin{tabular}{llll}
\hline \hline
Parameter & PPI model & noPPI model & R2012\\ 
 \hline \hline
 $m_1\sin(i)$ ($M_J$) & $0.92 \pm 0.04$ & $0.82 \pm 0.04$ & $0.85 \pm 0.05$ \\ 
 $m_2\sin(i)$ ($M_J$) & $0.85 \pm 0.03$ & $0.87 \pm 0.03$ & $0.82 \pm 0.07$ \\ 
 $a_1$ (AU) & $0.641 \pm 0.002$ & $0.643 \pm 0.004$ & $0.64 \pm 0.01$ \\
 $a_2$ (AU) & $1.017 \pm 0.005$ & $1.015 \pm 0.007$ & $1.02 \pm 0.02$\\
 $e_1$ & $0.17 \pm 0.03$ & $0.18 \pm 0.03$ & $0.17 \pm 0.03$\\
 $e_2$ & $0.10 \pm 0.04$ & $0.20 \pm 0.05$ & $0.16 \pm 0.1$\\
 $\omega_1$ & $178^{\circ +14^{\circ}}_{\ \ -17^{\circ}}$ & $142^{\circ} \pm 13^{\circ}$ & $143^{\circ} \pm 11^{\circ}$\\
 $\omega_2$ & $241^{\circ +72^{\circ}}_{\ \ -65^{\circ}}$ & $189^{\circ +10^{\circ}}_{\ \ -16^{\circ}}$ & $180^{\circ} \pm 26^{\circ}$\\
 $M_1$ & $91^{\circ +15^{\circ}}_{\ \ -18^{\circ}}$ & $124^{\circ} \pm 15^{\circ}$ & $129^{\circ} \pm 0.7^{\circ}$\\
 $M_2$ & $178^{\circ +73^{\circ}}_{\ \ -66^{\circ}}$ & $244^{\circ +12^{\circ}}_{\ \ -17^{\circ}}$ & $233^{\circ} \pm 0.9^{\circ}$\\
 $\gamma$ (m/s) & $3.76 \pm 0.71$ & $3.70 \pm 0.70$ & --- \\
 $J$ (m/s) & $2.90^{+0.78}_{-0.64}$ & $2.92^{+0.78}_{-0.62}$ & $2.49$\\
 \hline
\end{tabular}
\caption{Best-fit model parameters from our MCMC samples. Our PPI model includes planet-planet interactions, while our noPPI model excludes planet-planet interactions. The derived parameters from \R are also listed for ease of comparison.}
\label{tab:MCMC}
\end{table}

In the top panel of Figure~\ref{fig:MCMC} we plot the MAP estimate for our PPI model, along with 80 randomly drawn samples from our posterior distribution.
In the lower panel the residuals between our MAP estimate and the RV data are shown. 
As can be seen, our PPI model represents a good fit to the data. 

We also try adding two additional parameters to our PPI model to account for mutual inclination between the planets.
For this inclination model we add $i_{x,2} = 2\textrm{sin}(i_2/2)\textrm{cos}(\Omega_2)$ and $i_{y,2} = 2\textrm{sin}(i_2/2)\textrm{sin}(\Omega_2)$, where $i_2$ is the inclination of the outer planet (with respect to the inner planet and star) and $\Omega_2$ is the longitude of ascending node of the outer planet \citep{Pal2009}. 
The mutual inclination returned by this model is $i_2 = 30^{\circ +22^{\circ}}_{\ \ -37^{\circ}}$, being consistent with 0 and ruling out large mutual inclinations. 
Comparing this inclination model to our original PPI model using a posterior odds ratio (see Section~\ref{sec:MC}) yields a value of 0.4, indicating a slight preference over the inclination model.
Since our original PPI model is simpler and the model parameters between models are similar, we choose to stick with our PPI model for the remainder of our analysis. 


\subsection{Resonance Analysis}
Using our posterior samples found in Section~\ref{sec:Results}, we can assess the likelihood that the planets orbiting HD155358 are in 2:1 MMR. 
More specifically, we can draw samples from the posterior distribution and calculate what fraction of these systems have librating resonant angles. 
The resonant angles are calculated according to:
\begin{align*}
\begin{split}
\phi_1 &= (j - 1)\lambda_1 - j\lambda_2 + \omega_1 \\
\phi_2 &= (j - 1)\lambda_1 - j\lambda_2 + \omega_2 
\end{split}
\end{align*}
where $\lambda$ is the mean longitude, $\omega$ is the argument of periapsis, and $j$ is the order of the resonance ($j=2$ for this work).
Our aim is to use the fraction of systems with librating resonant angles as a measure of how likely the system is to be in MMR. 

For each set of parameters drawn from our posterior we simulate the corresponding system for 4000 years with a timestep equal to $P_1/50$. 
Over the course of a given simulation we generate 1000 equally spaced outputs for each resonant angle, and classify a resonant angle to be librating if the difference between the maximum and minimum is less than a threshold value $\kappa$. 
We set $\kappa = 7\pi/4$ since this allows for large resonant libration amplitudes while also ensuring that misclassification of librating/non-librating systems is low. 
Our results are insensitive to nearby values of $\kappa$ (e.g. $5\pi/3$, $9\pi/5$, etc.).
We check for librations around both 0 and $\pi$.

\begin{table}
\begin{tabular}{p{2cm}p{1.8cm}p{1.8cm}}
\hline \hline
 & PPI model & noPPI model \\\hline
$\phi_1$ librating & $98\% \pm 6\%$ & $1\% \pm 6\%$ \\
$\phi_2$ librating & $5\% \pm 6\%$ & $1\% \pm 6\%$ \\
 \hline \hline
\end{tabular}
\caption{
The percent of randomly drawn samples with $\phi_1$ and $\phi_2$ librating, for our PPI and noPPI models. 
Error bars calculated from simple Poisson statistics. 
}
\label{tab:res}
\end{table}

Our results are presented in Table~\ref{tab:res} for our PPI and noPPI models, for 300 samples randomly drawn from each posterior distribution. 
The error bars are calculated from simple poisson statistics.
For our PPI model, almost every drawn sample has $\phi_1$ librating, strongly suggesting that the system is in MMR. 
In contrast, our noPPI model shows no evidence of librating resonant angles.

Given our analysis, there is a high likelihood that the HD155358 system is in MMR, and is therefore likely to have formed via planet migration, as we show in the next section.

\section{Formation via Migration}
\label{sec:form}
\subsection{Methods}
We now use the results found in the previous section to explore the formation of HD155358 via migration.
We use a simple model by which the outer planet is migrated into 2:1 MMR using a parametric model, mimicking the presence of a protoplanetary disk.
Specifically, we introduce semi-major axis and eccentricity damping timescales, $\tau_a = a/\dot{a}$ and $\tau_e = e/\dot{e}$, respectively, where the two are related by a constant $K = \tau_a / \tau_e$. 
We refer to this class of convergent-migration models as ``CM models". 

We follow the implementation by \cite{Papa2000} and add an additional acceleration on each planet according to:
\begin{align*}
\begin{split}
\textbf{a}_{\rm mig} &=  -\frac{\textbf{v}}{\tau_a} \\
\textbf{a}_{\rm damp} &= -2\frac{(\textbf{v}\cdot\textbf{r})\textbf{r}}{r^2\tau_e},
\end{split}
\end{align*}
where $\textbf{v}$ is the velocity and $\textbf{r}$ is the position of the planet relative to the star. 
The migration prescriptions of \citet{Papa2000} were designed with Type-I migration in mind, yet our planets likely lie in the Type-II migration regime.
However, since we are interested primarily in seeing whether a simple migration model can reproduce the data, this model choice will suffice. 
More accurate, 3-D hydrodynamical simulations of Type-II planetary migration are significantly more complex and beyond the scope of this paper. 

For each simulation three parameters are independently varied -- the eccentricity damping timescale of the inner planet $\tau_{e_1}$, the eccentricity damping timescale of the outer planet, $\tau_{e_2}$, and the migration timescale of the outer planet, $\tau_{a_2}$.
The inner planet does not undergo migration by itself, i.e. $\tau_{a_1} = \infty$.
Initial values for $\tau_{a_2}$ are drawn from a uniform distribution in log-space between $10^{2.5}$ and $10^7$ years, while initial values for $K_1$ and $K_2$ are drawn between $10^{-1}$ and $10^3$.
Planet masses are drawn from our posterior distribution (Section~\ref{sec:Results}), all eccentricity and inclination values are initialized to zero, and the remaining orbital parameters are randomly drawn from uniform distributions. 

When planets migrate into MMR in the presence of a protoplanetary disk an equilibrium eccentricity $e_{eq}$ is reached, representing a stable balance between migration excitation and protoplanetary disk damping \citep[e.g.][]{Goldreich2014}. 
For each simulation we migrate the outer planet inwards for a time of $T_{\rm mig} = 5\tau_a$ years, after which $e_{eq}$ has reached a stable value. 
In cases where $T_{\rm mig} < 5000$ years we extend the simulation time to $T_{\rm mig} = 5000$ to ensure a stable solution. 
The migration parameters $\tau_{a_1}$, $\tau_{e_1}$ and $\tau_{e_2}$ are then logarithmically increased to $10^7$ years over the same amount of time ($T_{\rm mig}$), mimicking the dispersal of the protoplanetary disk. 
Simulations are independently checked to ensure that a) an equilibrium eccentricity is reached over $T_{\rm mig}$, and b) the disk dispersal is done adiabatically, ensuring that the resonance is not abruptly changed during the increase of $\tau_{a_1}$, $\tau_{e_1}$ and $\tau_{e_2}$.
Simulations that do not fulfill this criteria are discarded.

Once migration is turned off, a RV curve is generated from the simulation and fit to the original RV data using \emcee as before, but now with only five free parameters: 
$t_s$, a parameter allowing the rescaling of the orbital periods, 
$t_t$, a parameter accounting for an offset in time, 
$y_s$, a RV-stretch parameter to account for amplitude offsets (effectively rescaling the masses of both the planets and the star, while keeping the mass ratios and periods constant).
As in the original analysis, we also include  
$\gamma$, an offset parameter to account for any stellar drift along the line of sight, and 
$J$, a jitter parameter to account for underestimated measurement noise and intrinsic stellar noise in the RV data.
Each \emcee fit is run for an initial burnin phase of 200 steps with 100 walkers, after which the walkers are resampled near the best solution and run for 2000 steps. 


\subsection{Model Comparison}
\label{sec:MC}
In a Bayesian sense, each simulation represents a different model $M$ characterized by $\theta_{\rm model}$ = \{$\tau_{a_1}$, $K_1$, $K_2$\}, with each model having a unique MAP estimate for its parameters $\theta_{fit}$ = \{$t_s$, $t_t$, $y_s$, $\gamma$, $J$\}.
Note that our analysis thus incorporates the fact that many different models might explain the data equally well. 
A single model with parameters $\theta$ = $\{\tau_{a_1}$, $K_1$, $K_2$, $t_s$, $t_t$, $y_s$, $\gamma$, $J$\} on the other hand would make MCMC convergence more difficult. 

When multiple models can explain the data, the posterior odds ratio can be used to determine if there is any preference for one model over another. 
The posterior odds ratio, $P_{ij}$, is calculated according to \citep{Gregory2005}:
\begin{equation}
P_{ij} = \frac{p(M_i)}{p(M_j)} \frac{p(D|M_i)}{p(D|M_j)}, 
\label{eq:PO}
\end{equation}
where $p(M)$ is the prior odds for model $M$, and $p(D|M)$ is the marginal (or global) likelihood for model $M$.
The ratio of marginal likelihoods for two competing models is also known as Bayes' factor. 
In our case, the ratio of prior odds is unity since we have no prior model preference. 
Thus, the posterior odds ratio is equivalent to calculating Bayes' factor.

Formally, the marginal likelihood is calculated by marginalizing over the parameters $\theta$ of a model according to \citep{Gregory2005}:
\begin{equation}
p(D|M) = \int p(\theta|M)p(D|\theta,M)d\theta = \Lagr(M)
\label{eq:GL}
\end{equation}
where $p(D|\theta, M)$ is the likelihood and $p(\theta|M)$ is the prior. 
The marginal likelihood represents the probability of the data given the model. 

\begin{figure*}
\includegraphics[trim=4.4cm 7cm 4.5cm 0cm, width=\textwidth]{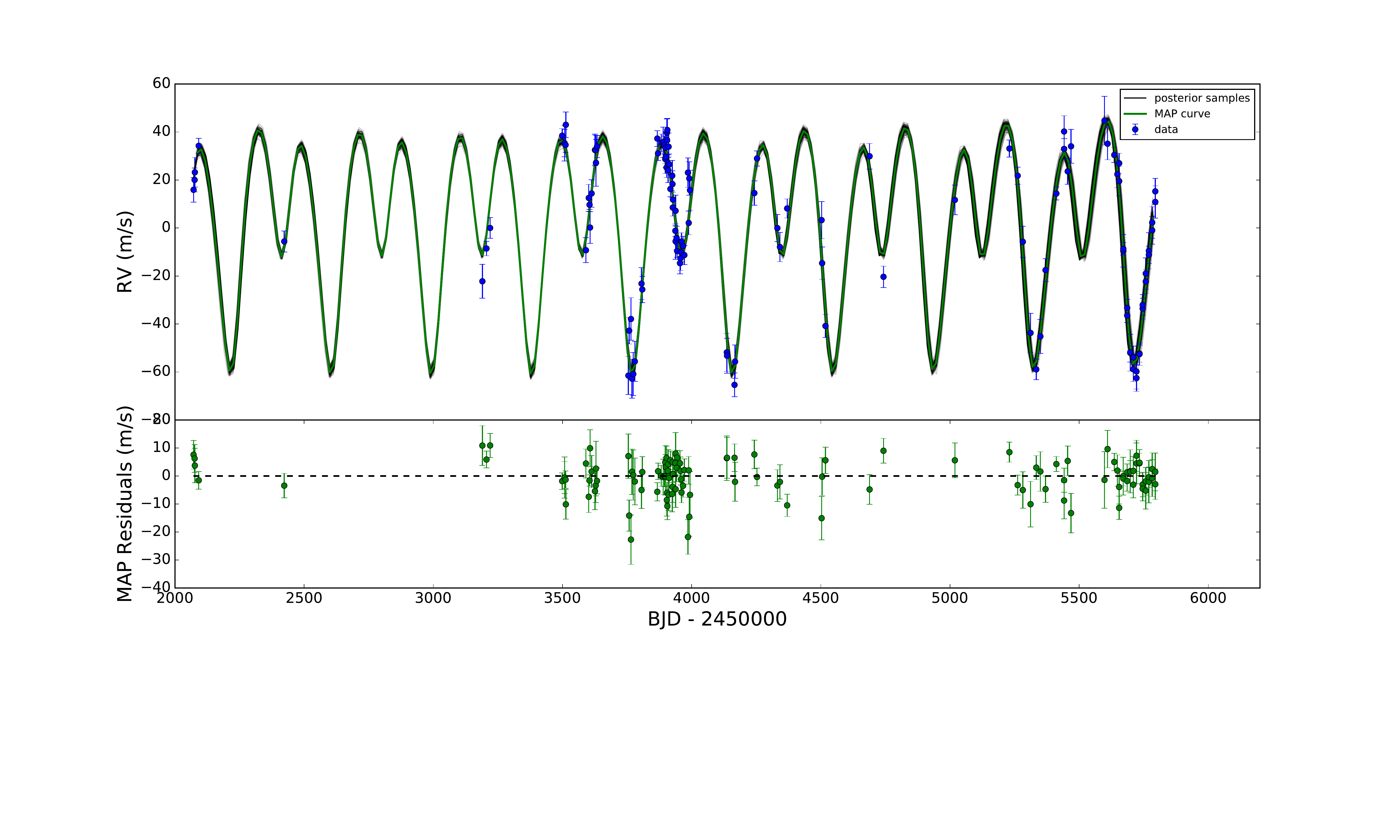}
\caption{RV data and $\theta_{\rm mig,best}$ fit.
Top panel shows the RV data points, MCMC MAP value (green), and 80 randomly plotted samples from the posterior (black). 
Bottom panel shows the residuals for the MAP fit to the RV data. 
 }
\label{fig:bestform}
\end{figure*}

\begin{figure}
\includegraphics[width=\columnwidth]{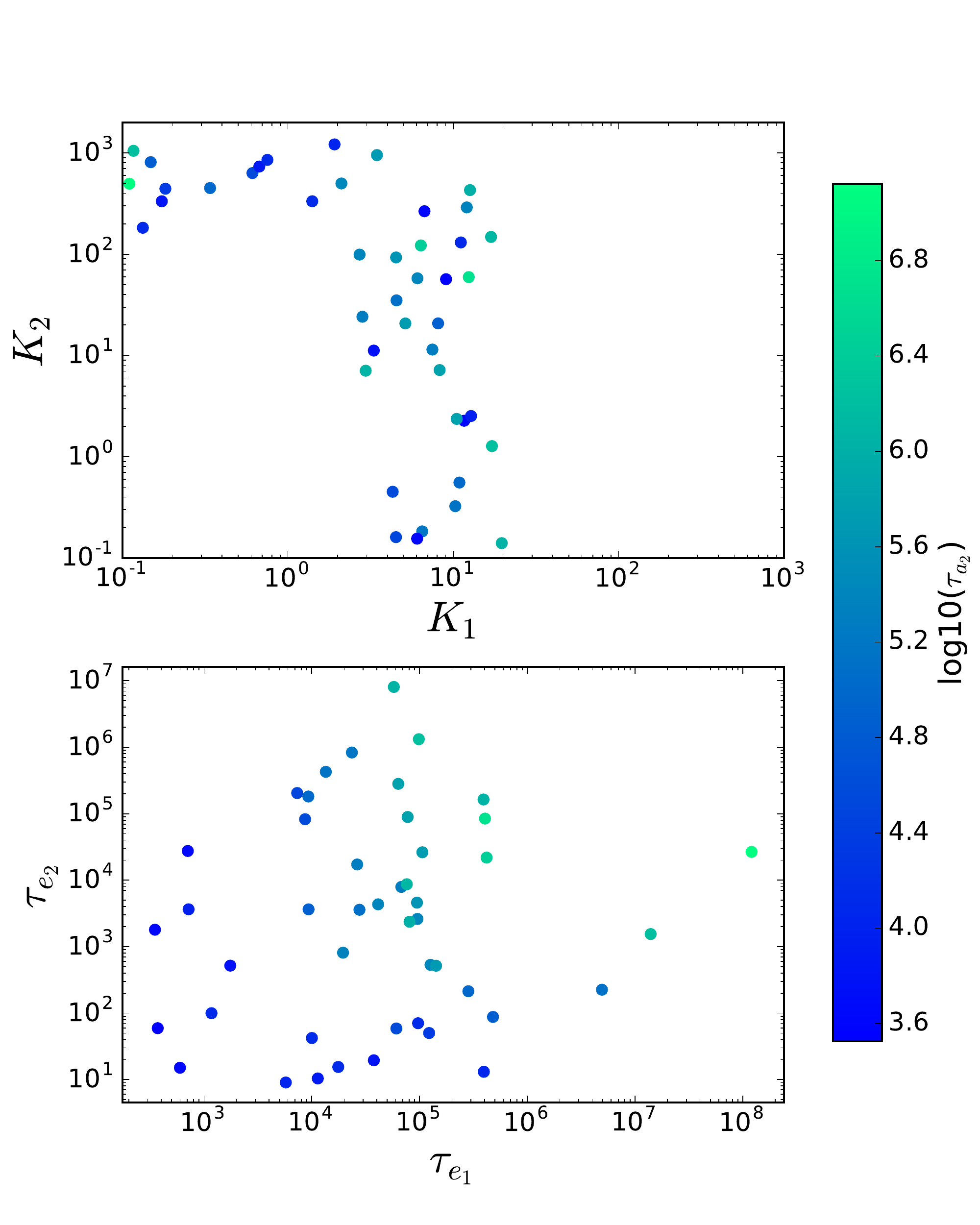}
\caption{The best 45 models $\theta_{\rm model}$ = \{$\tau_{a_1}$, $K_1$, $K_2$\} with Bayes' factors less than 100 when compared against the best model in our sample of 3000 simulations. 
Top panel presents the data in ($K_1$, $K_2$) space, and the parameters from our 3000 simulations were uniformly sampled from the region displayed. 
Bottom panel displays the same data in ($\tau_{e_2}$, $\tau_{e_2}$) space. 
Colour indicates the log value of $\tau_{a_1}$.
 }
\label{fig:goodones}
\end{figure}

When the marginal likelihood is normally distributed with flat priors, Eq.~\ref{eq:GL} can be approximated as \citep{Kass1995, Gregory2005}:
\begin{equation}
\Lagr(M) \approx \Lagr(\hat{\theta}) (2\pi)^{d/2} |\Sigma|^{1/2} \Delta \theta
\label{eq:Lapprox}
\end{equation}
where $\Lagr(\hat{\theta})$ is the the maximum likelihood, $d$ is the number of parameters in model $M$, $\Sigma$ is the covariance matrix of the posterior distribution, and $\Delta \theta = \prod_i^d 1/\delta \theta_i$ is the product of prior probabilities normalized by their lengths $\delta \theta$.
Since all models have the same flat priors for each parameter, Eq.~\ref{eq:PO} becomes:
\begin{equation}
P_{ij} = \frac{\Lagr_i(\hat{\theta}) |\Sigma_i|^{1/2}}{\Lagr_j(\hat{\theta}) |\Sigma_j|^{1/2}}
\label{eq:POapprox}
\end{equation}

Since \emcee returns both the samples and log-likelihoods at each step in the MCMC chain, it is straightforward to calculate the posterior odds ratio using Eq.~\ref{eq:POapprox}. 


\subsection{Results}
\label{sec:formresults}

Our best model from a sample of 3000 simulations is plotted in Figure~\ref{fig:bestform}, corresponding to $\theta_{\rm model, best}$ = ($\tau_{a_2}$, $K_1$, $K_2$) = (4000 years, 7, 265).
In the top panel we plot the MAP estimate along with 80 randomly drawn samples from our posterior distribution.
In the lower panel the residuals between our MAP estimate and the RV data are shown. 
As can be seen, our model represents a good fit to the data. 

Following \citet{Kass1995}, a posterior odds ratio lower than 100 indicates no decisive preference for one model over another, and in Figure~\ref{fig:goodones} we plot competing models which, when compared to $\theta_{model,best}$, yield a posterior odds ratio of less than 100. 
The top panel in Figure~\ref{fig:goodones} presents the valid models in ($K_1$, $K_2$, $\tau_{a_2}$) space, while the bottom panel presents them in ($\tau_{e_1}$, $\tau_{e_2}$, $\tau_{a_2}$) space. 
Although many different migration models are able to explain the data, regions of parameter space appear to be ruled out.
In the top panel of Figure~\ref{fig:goodones}, ($K_1 < 3$, $K_2 < 100$) and $K_1 > 20$ are two regions that appear to be ruled out by the data. 
In the bottom panel, positive correlation can be seen between $\tau_{a_2}$ and $\tau_{e_1} + \tau_{e_2}$.

\section{Stability}
\label{sec:stability}
\begin{figure*}
\includegraphics[trim=4.4cm 2cm 4.5cm 0cm, width=\textwidth]{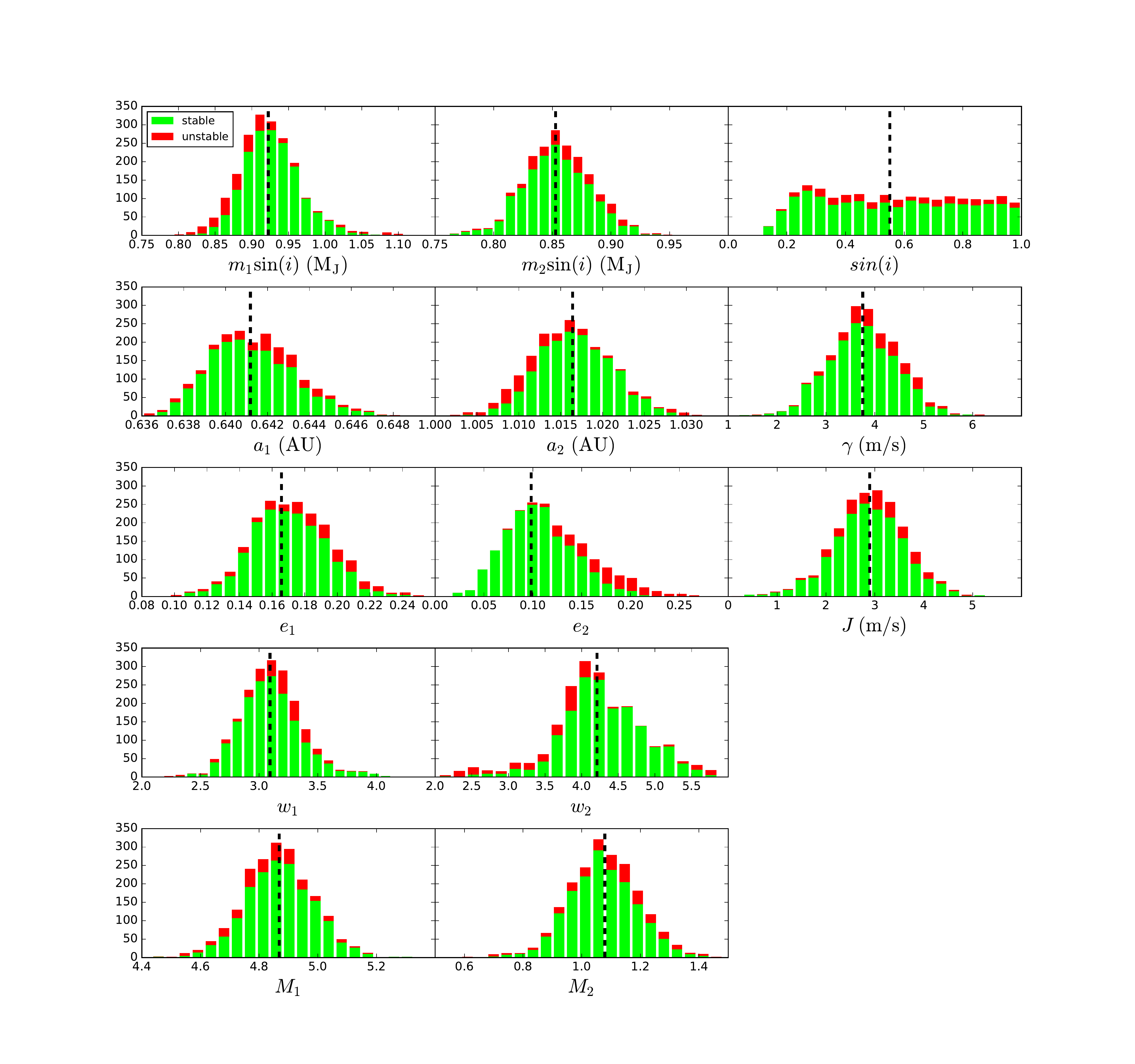}
\caption{
    Histograms of each parameter showing the distribution of 2000 stable (green) and unstable (red) systems drawn from our PPI model's posterior distribution and simulated for $10^9$ years.
The black vertical dashed line in each histogram represents our MAP estimate for that parameter. 
 }
\label{fig:stability}
\end{figure*}
Following \citet{Marshall2010} and \cite{Horner2011}, \R conducted a stability analysis by holding the inner planet constant at the best-fit values and varying the outer planet's orbital parameters randomly over a 3$\sigma$ range in $a$, $e$, $\omega$ and $M$. 
However, drawing parameters in this manner does not preserve their correlations, and it is possible to draw parameter combinations that are inconsistent with the RV data. 
\R also kept the planet masses fixed at their minimum $m\sin(i)$ values, precluding the possibility of further constraining the masses from the stability analysis. 

In a Bayesian framework, we can simply draw parameters from the posterior distribution and simulate them for a desired length of time. 
Parameter correlations are naturally preserved in the posterior distribution, and a probability of longterm stability can be easily found. 

In this study we draw 2000 random samples from our PPI model's posterior distribution and simulate the subsequent evolution for $10^9$ years using \whfast with a timestep of $P_1$/200. 
We also make use of the {\tt Simulation Archive} \citep{Rein2017} for stopping, restarting and analyzing simulations. 
The relative energy error for all stable simulations remains bounded at $< 10^{-7}$.

The results are presented in Figure~\ref{fig:stability} as a histogram for each parameter.
Stable and unstable systems over $10^9$ years are marked in green and red, respectively, while our MAP values from our PPI model is shown as vertical, black, dashed lines. 
$83\% \pm 2\%$ of our samples are stable over $10^9$ years and our MAP values land in highly stable regions, indicating that our PPI model is longterm stable. 

Unstable systems appear to be clustered in certain regions of parameter space.
As can be seen in Figure~\ref{fig:stability}, $m_1\textrm{sin}(i)<0.87$, $e_2>0.15$, $a_2<1.01$ AU and $\omega_2<4$ are regions where there are a significant fraction of unstable systems.
It is therefore unlikely that the true planet parameters lie in these regions. 

These unstable regions are consistent with the results found in \R, however, unlike \R our best-fit values are centred in highly stable regions (see Figures 9 and 10 in \R for a comparison). 
Although not shown, we find no clustering of unstable systems in ($m_1$, $m_2$) space, and thus the planet masses cannot be further constrained via stability analysis.  

\section{Discussion and Conclusion}
\label{sec:conc}
In this work we used three different model types to analyze the RV curve of HD155358 -- the PPI, noPPI and CM models. 
Our PPI (planet-planet interactions) and noPPI (no planet-planet interactions) models fit the RV curve to a model parameterized by the orbital elements of each planet (Section~\ref{sec:orb}), while our CM model was parameterized by migration parameters and fit this result to the RV curve (Section~\ref{sec:form}).
The best-fit parameters from our PPI and noPPI models are shown in Table~\ref{tab:MCMC}, while Table~\ref{tab:res} shows that there is a high likelihood that the planets of HD155358 are in MMR. 

The noPPI model is an approximation to the PPI model, and we have shown in Section~\ref{sec:Results} that planet-planet interactions are strong enough in the HD155358 system to yield differing conclusions about the resonant structure. 
More specifically, our PPI model predicts a high likelihood that the planets are in MMR, while our noPPI model predicts a low likelihood that the planets are in MMR. 
Since planets in MMR are highly likely to have formed via migration, by extension the PPI and noPPI models have different formation implications. 
Quantitatively, the orbital parameters returned by our PPI and noPPI models differ by up to $2\sigma$.

Our CM model has demonstrated that formation models of this style can be used to fit RV curves and constrain the initial conditions of a given system. 
One possibility for future work is to make the CM model more sophisticated, for example by modelling the planet-disk interactions in more realistic hydrodynamic simulations \citep{Rein2010}. 
With such a model, one could place further constraints on the formation of HD155358.

From a frequentist's point of view, the reduced $\chi^2$ values for our PPI, noPPI and best CM model are 1.4, 1.6 and 0.7 respectively, indicating that all three models are capable of fitting the data well.
However, our PPI model should be considered the only valid model when it comes to determining the true values of the observed system.
We also performed a stability analysis on our PPI model, finding that $83\% \pm 2\%$ of samples drawn from our posterior are stable over $10^9$ years.
We find regions of stable and unstable parameter space similar to \R, however unlike \R our best-fit model solution is centred in a stable region. 

\section*{Acknowledgments}
This research has been supported by the NSERC Discovery Grant RGPIN-2014-04553 and NSERC PGS-D grant.
We thank Daniel Tamayo for valuable discussions and Niels Oppermann for his Bayesian modelling advice. 

\bibliographystyle{apj}
\bibliography{HDbib.bib}

\end{document}